\documentclass[aps,prl,twocolumn,amsmath,amssymb,eqnarray,array]{revtex4}
\usepackage{graphicx}
\usepackage{dcolumn}
\usepackage{bm}
\begin{document}

\title{Low-frequency surface plasmon excitations in multicoaxial negative-index metamaterial cables}
\author{M. S. Kushwaha$^{1}$ and B. Djafari-Rouhani$^{2}$}
\address
{$^{1}$Department of Physics and Astronomy, Rice University, P.O. Box 1892, Houston, TX 77251, USA\\
$^{2}$IEMN, UMR-CNRS 8520, UFR de Physique, University of Science and Technology of Lille I, 59655 
Villeneuve d'Ascq Cedex, France}

\begin{abstract}

By using an elegant response function theory, which does not require matching of the messy boundary conditions, we investigate the surface plasmon excitations in the  multicoaxial cylindrical cables made up of negative-index metamaterials. The multicoaxial cables with {\em dispersive} metamaterial components exhibit rather richer (and
complex) plasmon spectrum with each interface supporting two modes: one TM and the other TE for (the integer order
of the Bessel function) $m \ne 0$. The cables with {\em nondispersive} metamaterial components bear a different
tale: they do not support simultaneously both TM and TE modes over the whole range of propagation vector. The
computed local and total density of states enable us to substantiate spatial positions of the modes in the spectrum. Such quasi-one dimensional systems as studied here should prove to be the milestones of the emerging optoelectronics
and telecommunications systems.\\

\noindent {\em OCIS codes:} 160.3918, 240.6680, 250.5403, 350.3618.

\end{abstract}
\date{\today}
\maketitle

\newpage

\section{INTRODUCTION}

The surface plasmon is a well-defined excitation that can exist on an interface that separates a  surface-wave active medium [with $\epsilon < 0$] from a surface-wave inactive [with $\epsilon > 0$] medium. It is characterized by the  electromagnetic fields that are localized at and decay exponentially away from the interfaces into the bounding media. In a conventional system, and in the simplest physical situation, an interface supports one and only one confined mode associated with either p-polarization or s-polarization. It may sound somewhat exaggeration, but it seems to be true that the plasmon excitation, in classical as well as in quantum structures and both theoretically and experimentally,
is the most exploited and best understood quasi-particle in condensed matter physics [1].

The recent research interest in surface plasmon optics has been invigorated by an experiment performed on the transmission of light through subwavelength holes in metal films [2]. This experiment has spurred numerous theoretical [3-6] as well experimental [7-11] works on similar structured surfaces: either perforated with holes, slits, dimples,
or decorated with grooves. It has been argued that resonant excitation of surface plasmons creates
huge electric fields at the surface that force the light through the holes, yielding very high transmission coefficients. The idea of tailoring the topography of a perfect conductor to support the surface waves resembling the behavior of the surface plasmons at optical frequencies was discussed in the context of a surface with an array of two-dimensional holes [6]. The experimental verification of this proposal has recently been reported [12-14] on the structured metamaterial surfaces which support surface plasmons at microwave frequencies. Because of their mimicking characteristics, these geometry-controlled surface waves were named {\em spoof} surface plasmons.

Talking of the negative-index metamaterials reminds us of another hot subject that has been drawing immense attention
of many research groups world-wide for the past few years. Proposed some four decades ago by Veselago [15], advocated
by Sir John Pendry in 2000 [16], and practically realized by Smith and coworkers in 2001 [17], an artificially
designed negative-index metamaterial, exhibiting simultaneously negative values of electrical permittivity $\epsilon(\omega)$ and magnetic permeability $\mu(\omega)$ and hence negative refractive index $n$, seems to have extended many basic notions related with the electromagnetism. It forms a left-handed medium, with the energy flow
${\bf E\times {\bf H}}$ being opposite to the direction of propagation vector, for which it has been argued that such
phenomena as Snell's law, Doppler effect, Cherenkov radiation are inverted. Metamaterials are also lately becoming
known as the basis of the proposals for designing cloaking device [18] and exotic fundamental phenomena such as anomalous refraction [19].


At the outset, it would be interesting to shed some light on how the plasma frequency is lowered in these metamaterials structured periodically with wire loops or coils. Some time ago, Pendry and coworkers [20] argued that any restoring force acting on the electrons will not only have to work against the rest mass of the electrons, but also against the self-inductance of such wire structures. This effect is of paramount importance in these wire structures. They went on arguing that the inductance of a thin wire diverges logarithmically with wire radius and confining the electrons to
thin wires enhances their effective mass by orders of magnitude. In other simpler format [7] one can, from Ohm's law ($j=\sigma E_{local}$), determine the effective conductivity for the inductive wire , and calculate an effective local dielectric function analogous to the Drude dielectric function, but with plasma frequency directly related to the inductance ($L$) of the unit cell (of length $l$) and wire spacing $d$ according to
$\omega_p=\sqrt{l/(d^2 L \epsilon_0 )}$. Thus reducing the wire radius enhances the inductance which thereby lowers
the plasma frequency of the system. Such estimates led them to predict the plasma frequency on the order of $\sim$
7 to 8 GHz.

In the present paper, we generalize our recent Green function (or response function) approach [21] to investigate
the propagation characteristics of surface plasmons in multicoaxial cables made up of right-handed medium (RHM)
[with $\epsilon >0$, $\mu >0$] and the left-handed medium (LHM) [with $\epsilon(\omega) <0$, $\mu(\omega) <0$] in
alternate shells starting from the innermost cable. In other words, we visualize a cylindrical analogue of a
one-dimensional planar superlattice structure bent round until two ends of each layer coincide to form a
multicoaxial cylindrical geometry. We prefer to name such a resultant structure as multicoaxial cables.

Such structures as conceived here may pave the way to some interesting effects in relation to, for example, the
optical science exploiting the cylindrical symmetry of the coaxial waveguides that make it possible to perform
all major functions of an optical fiber communication system in which the light is born, manipulated, and
transmitted without ever leaving the fiber environment, with precise control over the polarization rotation and
pulse broadening [22]. The cylindrical geometries are already known to have generated particular interest for
their usefulness not just as electromagnetic waveguides, but also as atom guides, where the guiding mechanism is governed mainly by the excited cavity modes. It is envisioned that the understanding of atom guides at such a
small scale would lead to much desirable advances in atom lithography, which in turn should facilitate atomic
physics research [23].

The rest of the paper is organized as follows. In Sec. II, we briefly focus on the strategy of the formalism
generalized to be applicable to the multicoaxial metamaterial cables. In Sec. III, we discuss several illustrative examples on the dispersion characteristics and the density of states of the relevant systems. Finally, we conclude
our findings in Sec. IV.

\section{THEORETICAL FRAMEWORK}

Recently, we have embarked on a systematic investigation of the surface plasmon excitations in the cylindrical coaxial shells made up of negative-index metamaterials interlaced with right-handed media within the framework of an exact Green-function (or response function) theory [21]. The knowledge of such excitations is considered to be fundamental to the understanding of the basics of the plasmonic optics. We consider the cross-section of these coaxial cables to be much larger than the de Broglie wavelength, so as to neglect the quantum size effects. We include the retardation effects but neglect, in general, the damping effects and hence ignore the absorption. In the state-of-the-art high quality systems, this is deemed to be quite a reasonable approximation [17]. Therefore we study the plasmon excitations in a neat and clean system comprised of multicoaxial metamaterial cables (MCMC), schematically shown in Fig. 1.

The formalism of the problem is a straightforward generalization of the theory presented in Ref. [21]. While it is always important to have a paper as much self-contained as possible, we think that reiterating all the necessary mathematical part from Ref. [21] would make it an undesirable repetition. The working strategy is systematically illustrated in Fig. 2, with all the necessary details. We call attention to the fact that Fig. 2 is a $2(n-1)\times 2(n-1)$ matrix, with all the elements outside the shaded regions being zero. Now `1' refers to the first perturbation specified by Eq. (3.8), `n' stands for the second perturbation specified by Eq. (3.15), and `2', `3', `4', ....., `(n-2)', `(n-1)' correspond to the third perturbation specified by Eq. (3.23) in Ref. [21] for the respective shells.
We would like to stress that our formalism is {\it not} a perturbative scheme, albeit we use the term `perturbation'
--- the term `perturbation', in fact, implies to the step-wise operation concerned with the problem.

It is also noteworthy that this theoretical framework knows no bound with respect to the number of media involved in
the system and/or their material characteristics. This implies that the general theory in Ref. [21] provides equal footing for the choice of all different LHM, all different RHM, or a combination thereof in the alternate shells. As such, an interested reader is only advised to focus on the final response functions given in Eqs. (8), (15), and (23)
in Ref. [21] in order to build the matrix depicted in Fig. 2.

\section{ILLUSTRATIVE EXAMPLES}

For illustrative examples, we have focused on the dispersive negative-index metamaterials characterized by $\epsilon(\omega) = 1 - \omega_p^2/\omega^2$, where $\omega_p$ is the plasma frequency (usually in the GHz range),
and $\mu(\omega) = 1 - F\omega^2/(\omega^2-\omega_0^2)$, where $F$ is, generally, chosen to be a constant factor
($<1$) and $\omega_0$ is the resonance frequency (also usually chosen to be in the GHz range). Since this is, to our knowledge, the first paper on such a complex system of multicoaxial metamaterial cables, we adhere to the backbone simplicity and think that the complexities, such as the choice of different metamaterials, different (conventional) dielectrics as spacers, different (irregular) radii, and geometrical defects would (and should) come later and hence
are deferred to a future publication. As such, our system is considered to be made up of the (same) RHM and the
(same) LHM in alternate shells starting from the innermost cable of radius $R_1$ and fix (the total number of media, including the outermost semi-infinite medium) $n=15$.

\begin{figure}[htbp]
\includegraphics*[width=7cm,height=6cm]{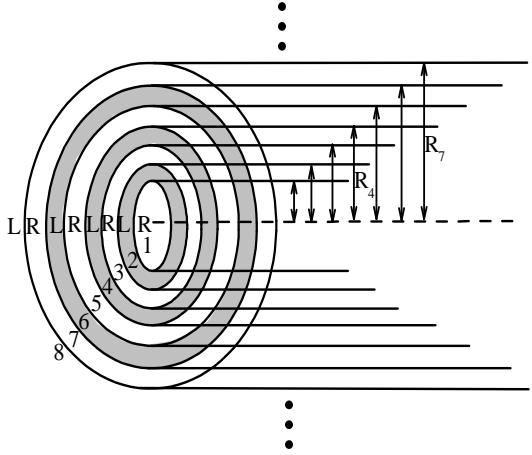}
\caption{Schematics of the multi-coaxial cables: the side view showing the alignments of the cylindrical cables of circular cross-sections of radii $R_{j+1} > R_j$, and $n$ media with $n-1$ interfaces. The innermost circle marked
as 1 refers to the innermost cable of radius $R_1$ enclosed by the consecutive ($n-1$) shells assumed to be numbered
as 2, 3, .... (n-2), (n-1) and cladded by an outermost semi-infinite medium $n$. Our exact general theory schematically
outlined in Fig. 2 allows one to consider the resultant system to be made up of negative-index (dispersive or nondispersive)  metamaterials interlaced with conventional dielectrics, metals, or semiconductors.}
\label{fig1}
\end{figure}


\begin{figure}[htbp]
\includegraphics*[width=7.5cm,height=7cm]{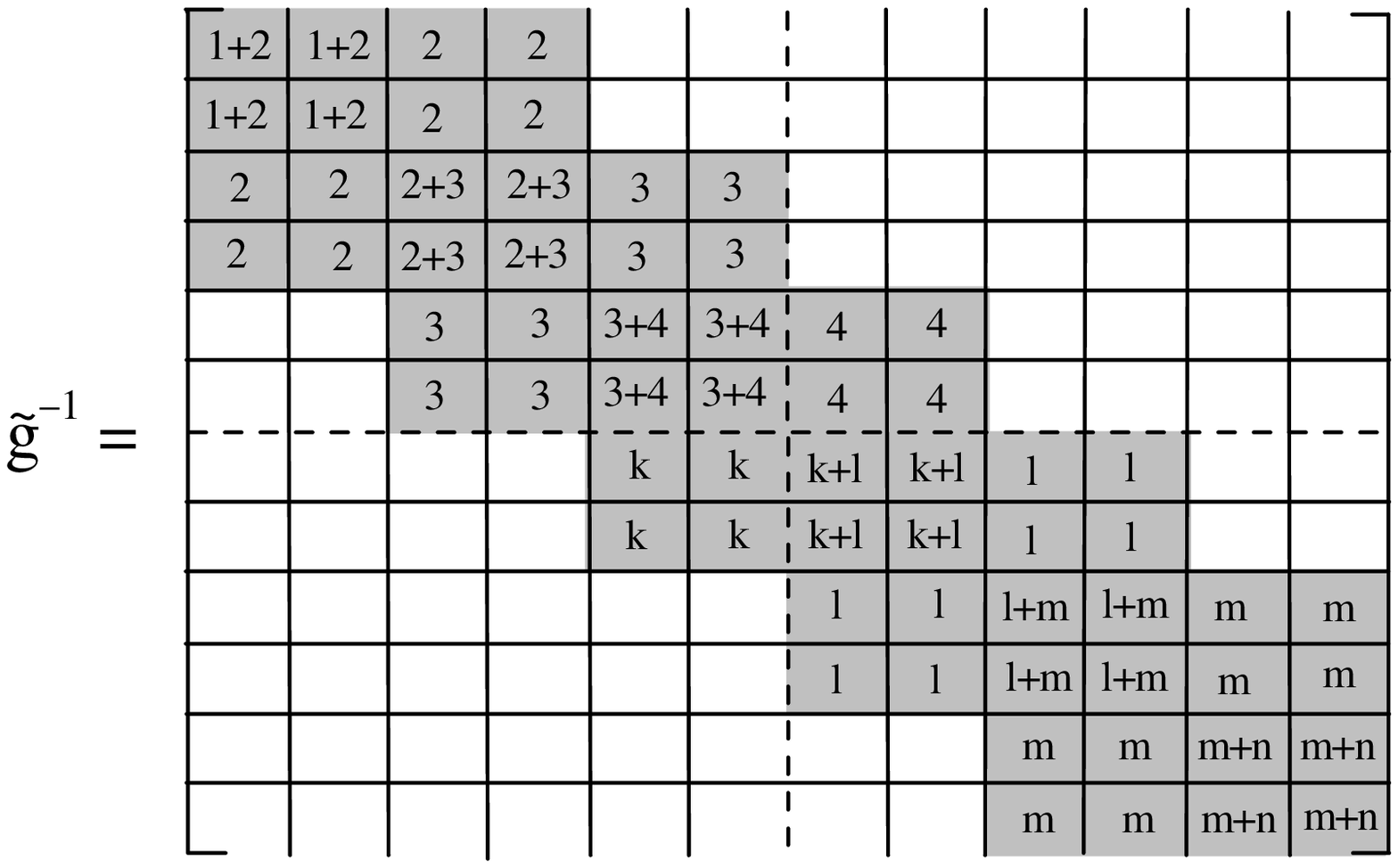}
\caption{A graphic representation of the complete formalism for the total inverse response function $\tilde{g}^{-1}(...)$ for the resultant system shown in a desired compact form. Here $n$ refers to the total number
of media comprising the MCMC system; with $m=n-1$ as the number of interfaces, $l=n-2$, and $k=n-3$ ..... etc. The plasma modes of the system are defined by $det [\tilde{g}^{-1}(...)]=0$.}
\label{fig2}
\end{figure}

Figure 3 illustrates the surface plasmon dispersion for a perfect multicoaxial cable system made up of a dispersive, negative-index metamaterials interlaced with conventional dielectrics (assumed to be vacuum) for $n=15$ and (the
integer order of the Bessel function) $m=0$. The plots are rendered in terms of the dimensionless frequency $\xi=\omega/\omega_p$ and the propagation vector $\zeta=c k/\omega_p$. The dashed line and curve marked as LL1 and LL2 refer, respectively, to the light lines in the vacuum and the metamaterial. The shaded area represents the region
within which both $\epsilon(\omega)$ and $\mu(\omega)$ are negative and disallows the existence of truly confined modes. The thick dark band of frequencies piled up near the resonance frequency $\omega_0=0.4 \omega_p$ is not unexpected. Since there are fourteen interfaces in the system, we logically expect fourteen branches each for the TM and TE modes
in the system. As we see, this is exactly the case, except for the fact that the lower group of seven TM branches
(which start from zero) have observed a resonance splitting due to the resonance frequency $\omega_0$ in the problem. The latter branches quickly become asymptotic to $\omega_0$. All the TM and TE confined modes above $\omega_0$ have their well-defined asymptotic limits exactly dictated, respectively, by
$\omega=\omega_p/\sqrt{\epsilon_1 +1}$ and $\omega=\omega_0/\sqrt{1-\frac{F}{\mu_1+1}}$.
A word of warning about the simultaneous existence of TM and TE modes: if we search the zeros of the determinant (see Fig. 2), as it is required, for {\em any} value of $n$ and $m$, we always obtain the simultaneous existence of all the TM and TE modes along with the resonance splittings as stated above. This is a {\em rule} as long as $m \ne 0$.
The only exception to this is the case of $m=0$ (and very small $n$). For instance, for $n=3$ and $m=0$, one has a
$4 \times 4$ determinant and it is possible to separate analytically the TM and TE modes. [We recall the well-known facts from the electrodynamics: the electrostatics (magnetostatics) claim ownership of the p-polarized (TM)
(s-polarized (TE)) fields.] However, even for these values of $n$ and $m$, if we search the zeros of the full determinant, without analytically decoupling the modes, we obtain both TM and TE modes together.

The interesting question that remains to be answered is: what is the advantage of investigating multicoaxial cables
over a few coaxial cables? The answer lies in comparing not only the growth mechanism but also the optimum response.
As to the fabrication, we understand that, for the reason of sensitivity, the growth conditions are more favorable
for the multicoaxial cables than for the single (or, a few coaxial) cables. This growth aspect is deemed to be just
similar to the growth mechanism of multiwalled carbon nanotubes. As regards the characteristic response of the multicoaxial cables, one can notice several differences as compared to that of a single (or a few coaxial) cable:
(i) the excitation spectrum becomes richer and complex, (ii) several waves separated by energy gaps can exist at
a given propagation vector, (iii) a single frequency allows several guided waves localized at different interfaces,
(iv) it should become feasible to describe the shells with varying effective parameters as required and realized in cloaking devices, and, most importantly, (v) the structure allows the coexistence of p-polarized and s-polarized
modes because the multicoaxial system is made up of metamaterials. We believe that these characteristics provide a
suitable platform for devising useful devices based on the surface plasmonic waves in the multicoaxial cables.

\begin{figure}[htbp]
\includegraphics*[width=7.5cm,height=8cm]{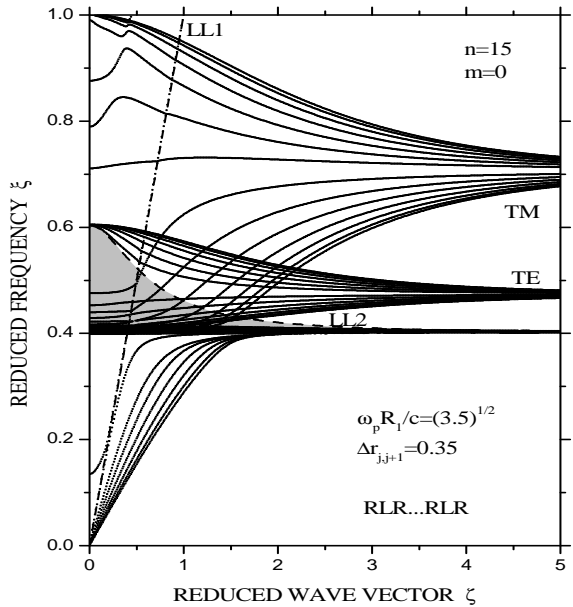}
\caption{Plasmon dispersion for a perfect multicoaxial cable system made up of a dispersive negative-index
metamaterial interlaced with conventional dielectrics for $N=15$ and $m=0$. The
dimensionless plasma frequency used in the computation is specified by $\omega_pR_1/c=\sqrt{3.5}$ and the
dimensionless thicknesses of the shells are defined by $\Delta r_{j,j+1}=0.35$. Dashed line and
curve marked as LL1 and LL2 refer, respectively, to the light lines in the vacuum and the metamaterial. The
(dark) thick band of frequencies is piled up at the characteristic resonance frequency ($\omega_0$). We call
attention to the resonance splitting of the lower TM confined modes due to the resonance
frequency ($\omega_0$) in the problem. The shaded area represents the region within which both $\epsilon(\omega)$
and $\mu(\omega)$ are negative and disallows the existence of truly confined modes. The system as a whole is
represented by RLR...RLR design.}
\label{fig3}
\end{figure}

Figure 4 shows the local density of states (LDOS) as a function of reduced frequency $\xi$ for the multicoaxial cable system discussed in Fig. 3, for the propagation vector $\zeta=1.0$.
The rest of the parameters used are the same as those in Fig. 3.
Notice that each of these interfaces is seen to share most of the peaks supposed to exist and reproduce most of the discernible modes at $\zeta =1.0$ (in Fig. 3). Of course, one has to take into consideration the degeneracy and the hodgepodge that persists near the resonance frequency $\xi=0.4$ in Fig. 3. Let us, for instance, look at the top panel:
the highest, second highest, third highest, fourth highest, fifth highest, sixth highest, seventh highest, and
eighth highest peaks lie, respectively, at $\xi=0.9490$, 0.9428, 0.9299, 0.9061, 0.8667, 0.8046, 0.7298, and 0.6232.
The highest peak (at $\xi=0.9490$) remains indiscernible at this scale.  Similarly, the lowest, second lowest, third lowest, fourth lowest, fifth lowest, sixth lowest, and seventh lowest peaks (below the resonance frequency) are seen
to lie, respectively, at $\xi=0.2877$, 0.2999, 0.3197, 0.3452, 0.3663, 0.3763, and 0.3967. All these peak positions exactly substantiate the modes at $\zeta=1.0$ in the spectrum in Fig. 3. One has to notice that, as the name
LDOS suggests, every interface has its own choice (with respect to the geometry and/or the material parameters) and there does not seem to be a rule that may dictate the modes' counting. This is, in a sense, different from the total
DOS where one obtains exactly the same number of peaks as the modes in the spectrum for a given value
of $\zeta$. Notice that the shorter-wavelength modes do not interact much with the neighboring ones and remain
spatially confined to the immediate vicinity of the respective interfaces. Such strongly localized modes are thus
easier to be observed in the experiments than their longer-wavelength counterparts.

\begin{figure}[htbp]
\includegraphics*[width=7.5cm,height=8cm]{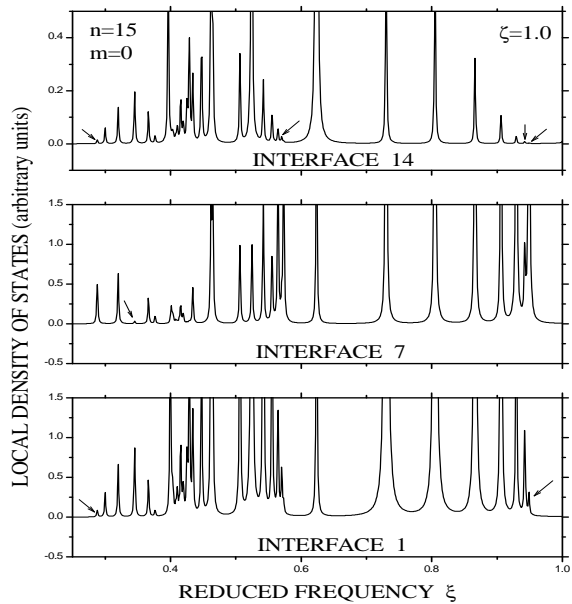}
\caption{Local density of states for the system discussed in Fig. 3 and for $n=15$, $m=0$, and $\zeta=1.0$. The bottom, middle, and top panel refer, respectively, to LDOS at interface 1, interface 7, and interface 14 in the system. The
rest of the parameters used are the same as in Fig. 3. The arrows in the panels indicate the relatively smaller (in
height) peaks.}
\label{fig4}
\end{figure}



Figure 5 depicts the surface plasmon dispersion (right panel) and total density of states (left panel) for a multicoaxial cable system made up of a nondispersive negative-index metamaterials interlaced with conventional dielectrics for $n=9$ and $m=0$. The plots are rendered in terms of the dimensionless frequency $\omega R_1/c$ and
the propagation vector $k R_1$. The dimensionless thicknesses of the shells are defined by $\Delta r_{j,j+1}=0.25$.
The material parameters are as listed inside the left panel.
Right panel: The dashed line refers to the light line in the vacuum. As expected, there are eight TM modes --- four
of them starting from the nonzero $k$ and the other four emerge from the light line.
The shaded area is the radiative region which encompasses radiative modes (not shown) towards the left of the light line. We notice in passing that the slope of these TM modes in the asymptotic limit is defined by $\omega/c k=0.7817$. An important issue remains to be answered: Why do we obtain only TM modes up to the asymptotic limit? The answer is
clearly provided by the analytical diagnosis. In order to answer this question, one has to look carefully at the analytical diagnoses presented in Sec. III.G in Ref. 21. To be brief, the answer lies in the fact that, for the
material parameters chosen here,  while Eq. (3.43) is fully satisfied, Eq. (3.47) or (3.48) is not. The former
condition justifies the existence of the TM modes and the latter rules out the occurrence of TE modes.
One remaining curiosity: What do these (almost) vertical lines, hanging downwards from the light line, indicated by arrows refer to? The succinct answer is that these are the ill-behaved TE modes which exist only in the long wavelength limit (LWL). It is interesting to note that if we interchange the values of $\epsilon_L$ and $\mu_L$,
(i.e., the parameters that define the nondispersive LHM), we obtain well-behaved TE modes and the ill-behaved TM modes.
The reason is simply that the aforesaid conditions that govern the nature of the modes in the asymptotic limit are then reversed. Left panel: The computation of the total density of states [plotted as a function of reduced frequency]
shows clearly eight peaks for the given value of $kR_1=21.5$. Starting from the lowest frequency, we observe that these peaks lie at $\omega R_1/c=14.69$, 14.99, 15.40, 16.20, 17.99, 18.53, 19.37, and 20.18. These peak positions exactly substantiate the frequencies of the TM modes in the right panel at $kR_1=21.5$.

The research efforts during the past few years reaffirm that the metamaterials are predominantly dispersive and lossy
materials. And yet, a considerable amount of research effort has focussed to explore interesting physical phenomena in the nondispersive metamaterials, particularly, in the context of photonic crystals in the recent years [24]. This leads us to infer that the results in Fig. 5 remain at least of fundamental interest.


\begin{figure}[htbp]
\includegraphics*[width=7.5cm,height=8cm]{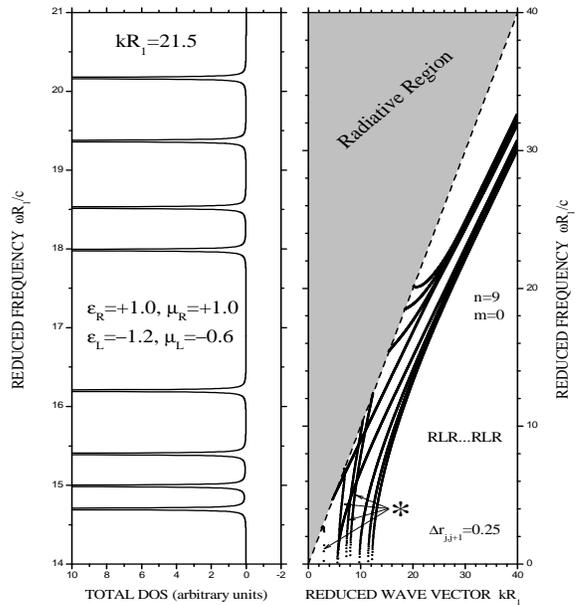}
\caption{Right panel: Plasmon dispersion for a perfect MCMC cable system made up of a non-dispersive negative-index metamaterial interlaced with conventional dielectrics for $n=9$ and $m=0$.
There are well-defined ($n-1$) TM modes in the system: upper four starting from the light-line and the
lower four from the nonzero propagation vector.
The dimensionless thicknesses of the shells are defined by $\Delta r_{j,j+1}=0.25$. The shaded region stands for the purely radiative modes (not shown). We call attention to the sharply downward modes, indicated by arrows, which are the ill-behaved TE modes in the LWL. Left panel: the total density of states as a function of reduced frequency $\omega R_1/c$ for the propagation wave vector $kR_1=21.5$.}
\label{fig5}
\end{figure}

\section{CONCLUDING REMARKS}

To conclude with, we estimate that if $\nu_p=\omega_p/2\pi=10$ GHz, the radius of the innermost cable is defined as $R_1=8.93$ mm for the parameter $\omega_pR_1/c=1.87$ used for the dispersive negative-index metamaterials (see Figs.
3 and 4). It is interesting to notice that this size scale is almost the same as the dimensions of the sample (the
lattice spacing $d=9.53$ mm and inner size of the tubes $a=6.96$ mm) used in the experiment in Ref. 12, which
verified the prediction of Pendry and coworkers [6] that, if textured on a subwavelength scale, even perfect
conductors can support the surface plasmon modes.

It is noteworthy that the frequency range of GHz is the most explored regime for the metamaterials realized from
split ring resonators and other similar geometries so far. Nevertheless, the recent fabrication processes can (and
do) allow to go to much higher frequency ranges approaching THz [10]. However, the appraisal of physical validity of
the effective parameters involved in the growth process is not so straightforward.

The surface plasmon modes predicted here should be observable in the inelastic electron (or light) scattering experiments. The EELS is already becoming known to be a powerful probe for studying the plasmon excitations in multiwalled carbon nanotubes. We trust that our methodology, as sketched in Fig. 2, will prove to be a powerful theoretical framework for studying further such plasma excitations in similar cable geometries such as multiwalled
carbon nanotubes.

\acknowledgments
M.S.K. gratefully acknowledges the hospitality of the UFR de Physique of the University of Science and Technology
of Lille 1, France, during the short visit in 2009. We sincerely thank Leonard Dobrzynski for many very fruitful discussions.


\newpage

\end{document}